\documentclass[twocolumn,
preprintnumbers,prl,
superscriptaddress
]{revtex4-1}

\usepackage{amsmath,amssymb}
\usepackage[usenames,dvipsnames,svgnames,table]{xcolor}
\usepackage[colorlinks,citecolor=DarkGreen,linkcolor=FireBrick,urlcolor=FireBrick,linktocpage]{hyperref}
\usepackage{graphicx}
\usepackage{times}
\usepackage{braket}
\usepackage{mathtools}
\usepackage{tikz}
\usetikzlibrary {decorations.pathmorphing}

\def\bC{\mathbb{C}}
\def\bZ{\mathbb{Z}}
\def\cD{\mathcal{D}}
\def\cE{\mathcal{E}}
\def\cF{\mathcal{F}}
\def\cH{\mathcal{H}}
\def\cX{\mathcal{X}}
\def\tr{\mathop{\mathrm{tr}}}
\def\A{U^H}
\def\B{U^{CZ}}
\def\elemdwall{\vcenter{\hbox{\begin{tikzpicture}[scale=.5]
\draw[decorate,decoration={zigzag,segment length=1mm, amplitude=.2mm}] (0,0)--(0,1);
\end{tikzpicture}}}}
\def\gwall{{\mid}}
\def\dwall {\ooalign{\hfil$\bullet$\hfil\cr\hfil$\elemdwall$\hfil\cr}}
\def\wwall{\ooalign{\hfil$\bullet$\hfil\cr\hfil$\elemdwall\elemdwall$\hfil\cr}}

\usepackage[subrefformat=parens]{subcaption}
\DeclareMathAlphabet{\mathdutchcal}{U}{dutchcal}{m}{n}
\SetMathAlphabet{\mathdutchcal}{bold}{U}{dutchcal}{b}{n}
\DeclareMathAlphabet{\mathdutchbcal}{U}{dutchcal}{b}{n}
\def\sD{\mathsf{D}}
\def\sg{\mathsf{g}}
\def\sh{\mathsf{h}}
\def\sX{\mathsf{X}}
\def\cg{\mathdutchcal{g}}

\long\def\revised#1{{#1}}

\begin{document}

\title{Non-invertible symmetries act locally by quantum operations}

\author{Masaki Okada}
\author{Yuji Tachikawa}
\affiliation{Kavli Institute for the Physics and Mathematics of the Universe (WPI), \\
 University of Tokyo,  Kashiwa, Chiba 277-8583, Japan}


\begin{abstract}
\emph{Non-invertible symmetries} of quantum field theories and many-body systems generalize the concept of symmetries by allowing non-invertible operations in addition to more ordinary invertible ones described by groups.
The aim of this paper is to point out that these non-invertible symmetries act on local operators by \emph{quantum operations}, 
i.e.~completely positive maps between density matrices,
which form a natural class of operations containing both unitary evolutions and measurements
and play an important role in quantum information theory.
This observation will be illustrated by the Kramers--Wannier duality of the one-dimensional quantum Ising chain,
which is a prototypical example of  non-invertible symmetry operations.
\end{abstract}

\maketitle

\section{Introduction}
In recent years, the concept of symmetries has received generalization in various directions in the theoretical study of quantum field theories and of condensed-matter systems.
One such generalization is to allow certain non-invertibility in the symmetry operations involved,
and the resulting structure is now known under the name of \emph{non-invertible symmetries} and is an active area of research.
Important examples of such operations have been known for decades before this fashionable name was coined, however,
and the most prototypical one is the Kramers--Wannier duality transformation $\sD$ of the Ising model.
This transformation commutes with the Hamiltonian at criticality, and as such plays a role analogous to ordinary symmetry operations.
That said, it does not quite square to one but rather satisfies \begin{equation}
\sD^2 = 1 + \sg
\end{equation} where $\sg$ is the $\bZ_2$ symmetry of the Ising model,
as explained in great detail e.g.~in various works \cite{Hauru:2015abi,Aasen:2016dop,Li:2023ani,Seiberg:2023cdc,Seiberg:2024gek}.

There is a conceptual question, however. 
Ordinary invertible symmetries are implemented by unitary transformations.
When we say we allow non-invertible symmetry operations, exactly which class of operations do we allow ourselves to use?

The aim of this letter is to answer this question, by pointing out that they act locally by \emph{quantum operations},
a notion prominent in quantum information theory and in the analysis of open quantum systems.
Here, quantum operations form a natural class of processes which can be performed on quantum systems,
including unitary evolutions, measurements, the introduction of ancillary degrees of freedom and tracing them out, and so on.
They are defined to be linear transformations which map density matrices to density matrices.
As such, they have to satisfy a certain positivity property, known under the name of \emph{complete positivity}.
Details on these notions can be found e.g.~in \cite{Nielsen_Chuang}.

Before proceeding, we note that this answer was actually already given in the context of two-dimensional continuum conformal  field theory treated using von Neumann algebras in a series of works by Bischoff and collaborators  \cite{Bischoff1,Bischoff2,Bischoff3,Bischoff4,Bischoff5},
but the discussions there unfortunately use mathematical notions unfamiliar to many of theoretical physicists. 
Here we would like to demonstrate that this observation holds much more generally in a language more understandable to us.
We also note that the answer was practically known in the case of  one-dimensional spin chains
to the authors of \cite{Lootens:2023wnl},
in which it was shown that any non-invertible symmetry operation can be realized by unitary quantum circuits and measurements, based on their formulation of dualities as matrix product operators \cite{Lootens:2021tet,Lootens:2022avn}.
As such, the authors do not claim much originality in this letter;
rather, the intention of the authors is to disseminate this important observation to a wider audience of physicists. 

The rest of the letter is organized as follows.
We first give a brief review of  quantum operations and non-invertible symmetries,
and then provide a very general argument that non-invertible symmetries act on local operators by quantum operations.
As the argument would be rather abstract, we will then illustrate the idea concretely using the case of the Kramers--Wannier duality transformation of the one-dimensional quantum Ising chain.
We will conclude with a number of remarks.

\section{Generalities}

\begin{figure*}
\begin{minipage}{0.6\linewidth}
\includegraphics[width=.85\textwidth]{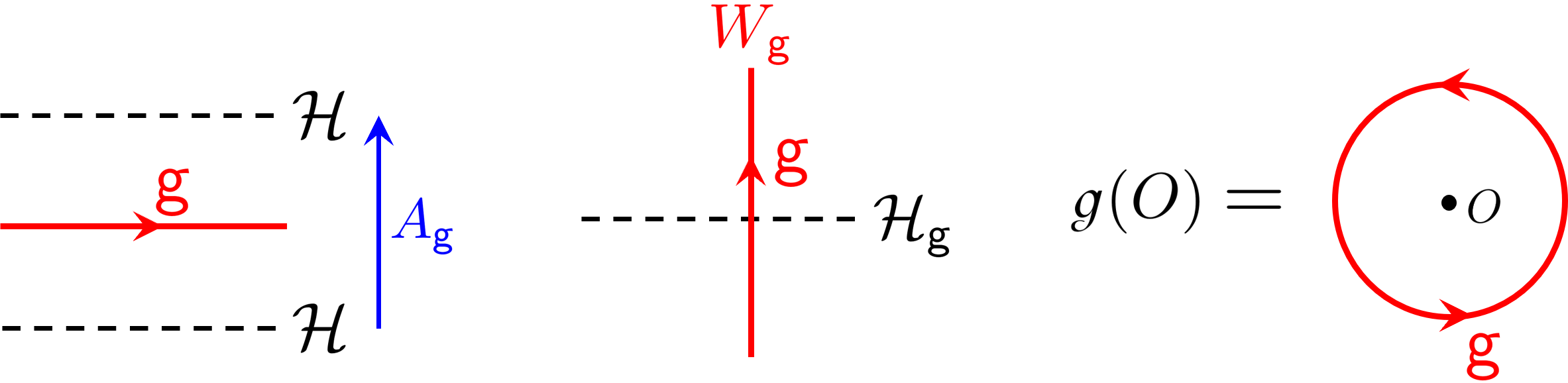}
\subcaption{\label{fig:walls1}}
\end{minipage}
\begin{minipage}{0.35\linewidth}
\includegraphics[width=.95\textwidth]{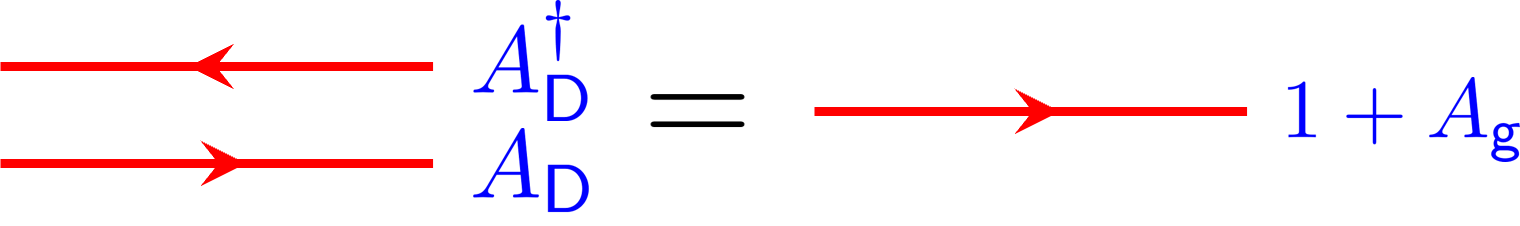}
\subcaption{\label{fig:walls2}}
\end{minipage}
\caption{\subref{fig:walls1} Symmetry actions as walls in space (shown horizontally) and time (shown vertically).
From left to right: as an action on the Hilbert space of states,
as a wall implementing a twisted boundary condition,
and as an operation on local operators.
\subref{fig:walls2} Fusion of two $\sD$ walls,
here represented as actions on the Hilbert space of states.
\label{fig:walls}}
\end{figure*}

\subsection{Quantum operations}
We start by recalling the concept of quantum operations \footnote{%
\revised{Depending on the subcommunity within physics,
what we call ``quantum operations'' here are often also called ``quantum channels''.
There are also communities where the terminology ``quantum channels'' refer to ``quantum operations'' in the sense used here with the additional constraint that it preserves the trace.
Here we stuck with the terminology of \cite{Nielsen_Chuang}.}}.
We will be brief; for details, the readers are referred to the standard textbooks, such as \cite{Nielsen_Chuang}.
Given a density matrix $\rho$ describing a quantum system with the Hilbert space $\cH$, 
we consider an operation of the form  $\rho \mapsto \rho'=\cE(\rho)$.
First, to preserve the statistical interpretation of density matrices, we require that $\cE(s\rho_1 +(1-s)\rho_2)=s\cE(\rho_1)+(1-s)\cE(\rho_2)$.
This motivates us to define $\cE$ as a linear map on the space of operators on $\cH$. 
Second, density matrices $\rho$ have positive eigenvalues; such operators are called positive operators.
Then, $\cE$ should map positive operators to positive operators; such maps are called positive maps.
Now, the operation $\cE$ also acts naturally on operators on the enlarged Hilbert space $\cH\otimes \bC^N$,
where $\bC^N$ describes  decoupled auxiliary degrees of freedom.
We require that $\cE$ should act on operators on $\cH\otimes \bC^N$ positively for all $N$;
such an $\cE$ is called a completely positive map.
Quantum operations are then defined to be completely positive maps on operators on a Hilbert space $\cH$;
these operations include both unitary evolutions and measurements.

A natural subclass of quantum operations is defined by the condition that $\tr \cE(\rho) = \tr \rho$,
meaning that $\cE$ preserves the total probability and describes the entire outcome of an operation rather than a specific subset. 
Such a completely positive map is called trace-preserving.

So far we used the Schr\"odinger picture where the operation $\cE$ acts on the density matrices.
We can instead use the Heisenberg picture and think of the operation as acting on observables as $O\mapsto \revised{\cF}(O)$, 
by postulating $\tr \revised{\cF}(O) \rho = \tr O\cE(\rho)$.
This $\revised{\cF}$ is also completely positive. When $\cE$ is trace-preserving, $\revised{\cF}$  satisfies the condition $\revised{\cF}(1)=1$ and is called a unital map.
In the rest of the paper we always use the Heisenberg picture.
For convenience, we will slightly weaken the unital condition to allow a scalar-multiple: $\revised{\cF}(1)\propto 1$.

Any quantum operation $\revised{\cF}$ on operators on $\cH$ is known to be represented as the combination of
(i)  introduction of an ancillary Hilbert space $\cH'$,
(ii) a unitary evolution on the combined system $\cH\otimes \cH'$, and
(iii) removal of the ancillary Hilbert space $\cH'$ by a measurement or by  the partial trace.
When we regard $\cH'$ as the environment, this can be interpreted as describing the noise introduced by the environment;
when we regard $\cH'$ as the measurement device, this procedure gives the effect of a measurement onto the target system.

More generally, we can use an entirely different Hilbert space $\cH''$, a linear map $V:\cH\to \cH''$, and a representation $\pi$ of the algebra of operators on $\cH''$. 
Then the combination $\revised{\cF}(O):=V^\dagger \pi(O) V$ is a quantum operation,
and this form is known as the Stinespring representation of $\revised{\cF}$.

\subsection{Non-invertible symmetries}

Let us next review the concept of non-invertible symmetries. 
Again we will be very brief; more details can be found in various lecture notes, e.g.~\cite{Schafer-Nameki:2023jdn,Brennan:2023mmt,Bhardwaj:2023kri,Luo:2023ive,Shao:2023gho,Carqueville:2023jhb}.
An ordinary symmetry operation $\sg$ of a quantum system described by a Hilbert space $\cH$ is given by a unitary operator $A_\sg$ on it.
When $\sg$ acts locally in a many-body or quantum field theory setting, we can consider a wall $W_\sg$ in space, 
across which we twist the system by $\sg$.
For a system on a circle with one such wall, this represents a twisted boundary condition.
The resulting Hilbert space is the twisted one $\cH_\sg$.
The action of $A_\sg$ can also be considered as the insertion of a wall in spacetime, spread along the spatial direction.
Then the action $\cg(O)$ of $\sg$ on an operator $O$ supported in a region of the space is given by wrapping the wall $W_\sg$ around the operator, which is equivalent to $A_\sg^\dagger O A_\sg$.
See Fig.~\ref{fig:walls1} for an illustration of the discussions so far.
For a more detailed explanation for the graphical detail, see the section 1 of Supplementary Material.

One important feature of these walls is that they can be freely moved in space and time as long as they do not hit other operators.
Another feature is that they fuse according to the group law, $W_{\sg\sh}=W_\sg W_\sh$, and as such they are invertible: $W_\sg W_{\sg^{-1}}=1$.
These two features are independent, and non-invertible symmetries are obtained by dropping the second property.

A prototypical example of non-invertible symmetry operations is the Kramers--Wannier duality transformation of the Ising model.
In the language of the one-dimensional quantum Ising chain, this transformation exchanges the Hamiltonians 
$\sum_i (a X_i + b Z_{i} Z_{i+1})$ and $\sum_{i'} ( b X_{i'} + a Z_{i'} Z_{i'+1})$.
The operator $A_\sD$ implementing this exchange on a closed chain however does not square to identity,
and is known to satisfy $A_\sD^\dagger A_\sD=1+A_\sg$, where $A_\sg=\prod_i X_i$ is the $\bZ_2$ symmetry operation.
As we will review in more detail below,
this duality transformation can be implemented locally on the Ising chain.
Correspondingly, we can consider the duality wall $W_\sD$ which fuses according to the rule $W_{\overline{\sD}}W_\sD = 1+ W_\sg$.
See Fig.~\ref{fig:walls2} for an illustration of the ideas described here.
Note that the action $\cD(O)$ of a non-invertible symmetry $\sD$ on an operator $O$, implemented by wrapping the wall $W_\sD$ around the operator, is no longer equivalent to $A_\sD^\dagger O A_\sD$.

\subsection{Non-invertible symmetries act locally by quantum operations}

\begin{figure}
\includegraphics[width=.45\textwidth]{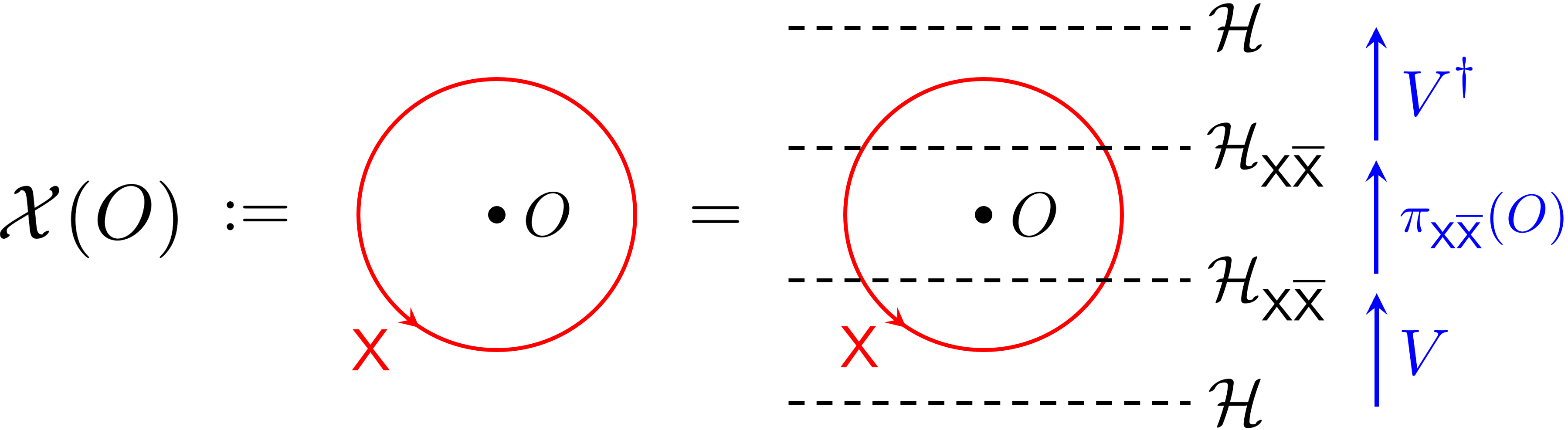}
\caption{Action of a non-invertible symmetry $\sX$ on a local operator $O$,
making it manifest that it is a Stinespring representation of a completely positive map. \label{fig:main} }
\end{figure}

\revised{We are now led to the following natural question. 
The action  $O\mapsto \cD(O)$ of a non-invertible symmetry $\cD$  is clearly a linear map. 
Does it have any other general properties? 
Our answer is that it is a quantum operation.}
The demonstration is surprisingly simple and general, see Fig.~\ref{fig:main}.
Consider the action of a non-invertible symmetry operation $\cX$ on an operator $O$ acting on $\cH$.
As shown in the figure,
we first go from the original Hilbert space $\cH$ to the Hilbert space $\cH_{\sX\overline{\sX}}$ with a wall $W_\sX$ and another wall $W_{\overline{\sX}}$ inserted, by a linear map $V$.
We then act by the operator $O$ on $\cH_{\sX\overline{\sX}}$, which is denoted by $\pi_{\sX\overline{\sX}}(O)$.
We now come back to the original Hilbert space $\cH$ by $V^\dagger$.
Then we find \begin{equation}
\cX(O) = V^\dagger \pi_{\sX\overline{\sX}}(O) V,
\end{equation} which is precisely a Stinespring representation of a completely positive map.
This argument is very general, and does not assume whether the theory is defined in continuum or on a lattice.
It does not assume that we have tensor-product Hilbert spaces either.
The argument is admittedly  very abstract, however.
We will now make it more concrete by studying the case of the Kramers--Wannier duality wall explicitly.

\section{Kramers--Wannier duality}

We consider a one-dimensional spin chain, with each site  $\bullet$ described by a qubit $\ket{0}$, $\ket{1}$ spanning a Hilbert space $\cH(\bullet)=\bC^2$.
Let $X$, $Y$, $Z$ denote the Pauli matrices acting on the qubit.

\revised{The way the $\bZ_2$ symmetry acts on the spin chain is well-known.
We provide a detailed explanation of how it is represented graphically in the Supplementary Material,
to which we refer readers unfamiliar with this graphical technique.}
Let us then directly discuss the wall $\elemdwall$ implementing the Kramers--Wannier duality $\sD$, which was already studied in great detail in \cite{Hauru:2015abi,Aasen:2016dop,Li:2023ani,Seiberg:2023cdc,Seiberg:2024gek}.
We take the convention that the wall sits on top of a site, as in $\cdots\bullet\dwall \bullet\cdots$.
Then the motion of the wall to the right by one unit is given by 
\begin{equation}
\hspace{-2pt}
\cH(\cdots \dwall _i \bullet_{i+1}\cdots) \xrightarrow{R_{\sD,i+1}:= \A_{i+1} \B_{i,i+1}} \cH(\cdots \bullet_i \dwall _{i+1}\cdots) ,
\end{equation}
where $\B$ acts on two qubits by the controlled-$Z$ gate
\begin{equation}
\bra{s't'}\B\ket{st}=\delta^{s'}_s \delta^{t'}_t (-1)^{st},
\end{equation}
and $\A$ acts on a single qubit by the Hadamard gate
\begin{equation}
\bra{t'}\A\ket{t}=\sqrt{2}^{-1}(-1)^{t't}.
\end{equation}
Two walls are introduced by the following operation:
\begin{widetext}
\begin{align}
\begin{array}{cccc}
V_{\sD\overline{\sD},i} \colon &
\cH(\cdots  \bullet_i \bullet_{i+1} \cdots) &
 \xrightarrow{\otimes \ket{0_a}} \cH(\cdots \bullet_i  \wwall_a \bullet_{i+1}\cdots) 
 \xrightarrow{R_{\sD, a}^\dagger} &
 \cH(\cdots \dwall_i  \dwall_a \bullet_{i+1}\cdots),\\
 &\rotatebox{90}{$\in$} &&\rotatebox{90}{$\in$} \\[-1ex]
&  \ket{\cdots s_i s_{i+1} \cdots } & \xmapsto{\phantom{very long very long indeed very very long}} &
\sqrt{2}^{\frac{1}{2}} R_{\sD, a}^\dagger \ket{\cdots s_i 0_a s_{i+1} \cdots }
\end{array}
\end{align}
\end{widetext}
i.e.\ tensoring the state $\ket{0}_a\in\cH(\wwall_a)$ of the ancillary qubit and detaching one of the walls to the left \footnote{
There is a subtlety in the notation \unexpanded{$\wwall_a$} of two $\sD$ walls on the same site,
in that separating the wall to the left is given by $R^\dagger_{\sD}$ but separating the wall to the right is not simply given by $R_{\sD}$.
The latter operation needs to be given by $R_{\sD,a} R_{\sD,i+1} V_{\sD\overline{\sD}, i}$, i.e.~
we always have to use $V_{\sD\overline{\sD}}$ when introducing or canceling two $\sD$ walls.
See \cite{Hauru:2015abi,Aasen:2016dop,Li:2023ani,Seiberg:2023cdc,Seiberg:2024gek} for other conventions.}.

Let $O$ be an operator supported well within the sites from $j$ to $k$, with $j<k$. 
Our definition leads to the local action of $\sD$ given by \begin{multline}
\cD(O)=V_{\sD\overline{\sD},k}^\dagger
(R_{\sD,k}
R_{\sD,k-1}
\cdots
R_{\sD,j}
) \times \\
O
(R_{\sD,j}^\dagger
\cdots
R_{\sD,k-1}^\dagger
R_{\sD,k}^\dagger)
V_{\sD\overline{\sD},k},
\label{eq:D-action-on-O}
\end{multline}
\revised{which manifestly is in a Stinespring representation,
See Fig.~\ref{fig:Dwalls}.
It can also be readily converted into a Kraus representation,
as we concretely show in the Supplementary Materials.}

\begin{figure}[t]
\includegraphics[width=.25\textwidth]{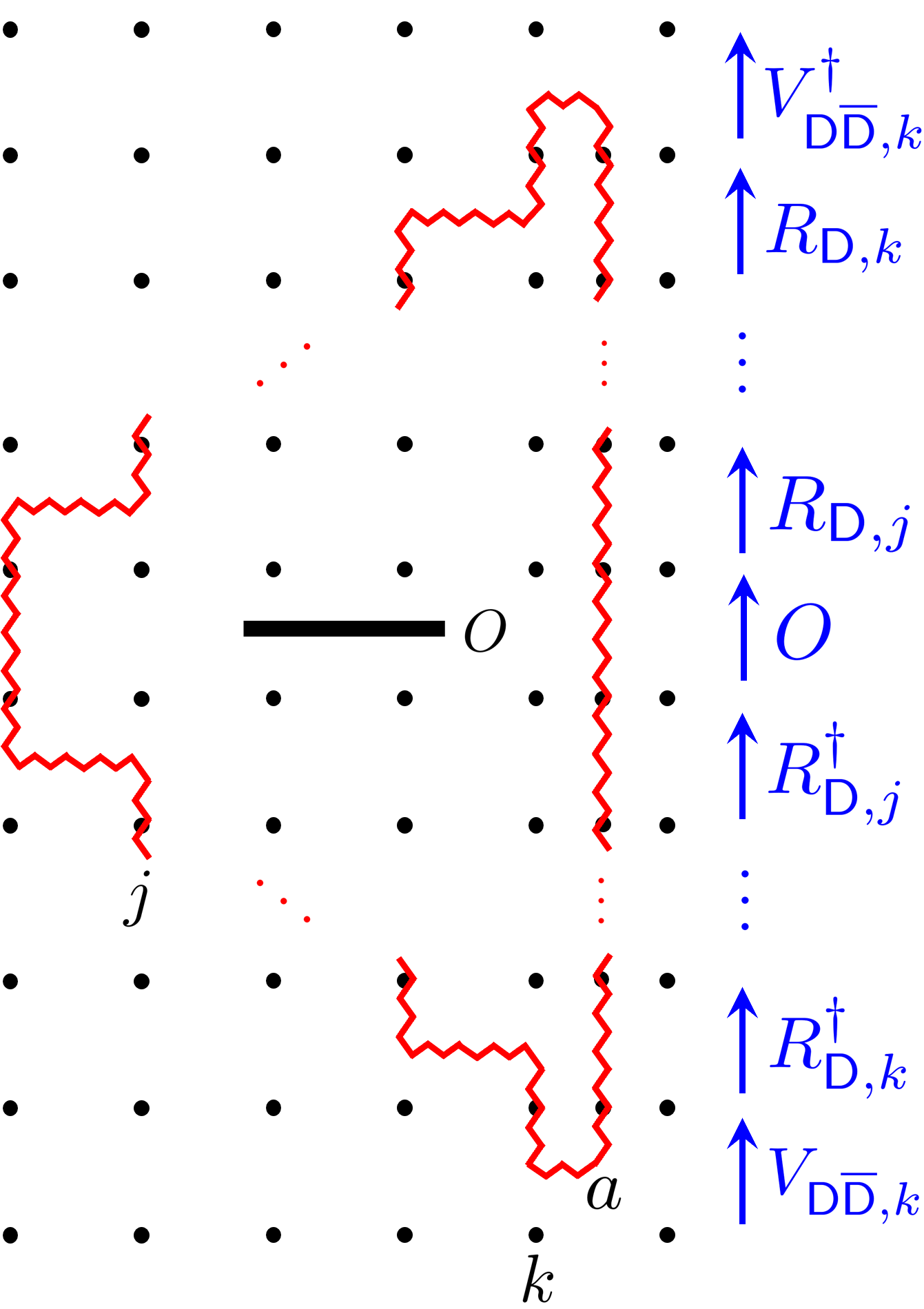}
\caption{\revised{Action of a non-invertible symmetry $\sD$ on a local operator $O$ implemented by the walls in the model of a one-dimensional spin chain. \label{fig:Dwalls}}}
\end{figure}

\revised{Let us study the action $\cD(O)$ in an example.} Let 
\begin{equation}
H=\sum_{i=j'}^{k'-1} (a X_{i}+ b Z_{i}Z_{i+1})+a X_{k'}
\end{equation}
be the standard Hamiltonian of the Ising chain restricted to the sites $j'$ to $k'$. 
Assuming  $j\leq j'$ and $k'\leq k$, we can use the formula above to compute the local action of $\sD$, and we find \begin{equation}
\cD(H) = \sqrt{2} \left[ \sum_{i=j'}^{k'-1} (a Z_{i-1}Z_{i}+ b X_{i})+a Z_{k'-1}Z_{k'} \right],
\end{equation}
as expected for a local action of Kramers--Wannier duality.
With some efforts, we can show that \begin{equation}
\mathcal{R}(\cD^2(O))=O+\cg(O) \label{foo}
\end{equation}
in general, where $\mathcal{R}$ is the operation to move an operator one unit to the right \footnote{
For example, we can easily check $\cD(X_i) = \sqrt{2}Z_{i-1}Z_i$ and $\cD(\sqrt{2}Z_{i-1}Z_i) = 2 X_{i-1}$ from \eqref{eq:D-action-on-O}.
If we prefer not to have the appearance of the lattice translation here,
we can consider $\cD'(O)$, which introduces two $\sD$ walls at a site on the left of $j-1$, moves one wall to the right beyond $k+1$, acts $O$, and reverses the wall motion.
Then $\cD'(\cD(O)) = O + \cg(O)$ follows.
Another option is to use a convention that distinguishes spins and dual spins \cite{Aasen:2016dop,Li:2023ani}.
}. For example, for $j \leq i \leq k$, we can easily check \begin{equation}
\cD(Z_i) =0,
\end{equation}
which is compatible with \eqref{foo}, as $\cg(Z_i)=-Z_i$.

Before proceeding, we note that the implementation of Kramers--Wannier duality operation and broader classes of non-invertible symmetries by quantum circuits and measurements was discussed in many other places, e.g.\ \cite{Fukusumi:2020irh,Ashkenazi:2021ieg,Tantivasadakarn:2021vel,Lootens:2021tet,Aasen:2022cdu,Bravyi:2022zcw,Tantivasadakarn:2022hgp,Lootens:2022avn,Lootens:2023wnl,Fechisin:2023dkj,Okuda:2024jzh}.

\section{Concluding Remarks}

In this letter we gave a general argument that non-invertible symmetries act on locally-supported operators by quantum operations.
We also illustrated this observation in the concrete case of the Kramers--Wannier duality of the one-dimensional quantum Ising chain,
where we can explicitly see the introduction of an ancillary qubit in a specific state,
the action of unitary operator on the combined system,
and then the removal of the ancillary qubit.

The authors admit that what they have here is simply a curious observation, 
and that it was already noted several years ago in the context of algebraic quantum field theory in \cite{Bischoff1,Bischoff2,Bischoff3,Bischoff4,Bischoff5},
and that it was essentially understood in the context of spin chains in \cite{Lootens:2023wnl,Lootens:2021tet,Lootens:2022avn}.
\revised{The authors think} that it was still worthwhile to spend a few pages to explain this observation in a language understandable to more theoretical physicists,
especially because this observation might open a fruitful flow of ideas between two actively researched areas.

For example, we can ask what the Petz recovery map on the quantum operation side corresponds to, if any, on the side of non-invertible symmetries. 
Is it related to existence of the dual $\overline{\sX}$ for any non-invertible operation $\sX$ such that $\overline{\sX}\sX$ contains the identity?
Another natural question is to ask if the property of a general quantum system, not necessarily associated to quantum field theory or quantum many-body systems,
would be affected or constrained in any way by the existence of a non-invertible symmetry structure.
For example, the  existence of an ordinary symmetry often leads to degeneracy in the spectrum.
Can we say anything interesting about the spectrum of a Hamiltonian,
if it is assumed to be invariant under two quantum operations $\cD$ and $\cg$ satisfying $\cD^2=1+\cg$?
\revised{Another possible direction is to consider not just 0-form non-invertible symmetries, as in this letter, but other higher-form non-invertible symmetries.}
The authors hope to come back to these questions in the future, and
the authors would also welcome other researchers to do so.

\section{Acknowledgments} 
The authors thank Kotaro Kawasumi for asking exactly which class of operations non-invertible symmetries are; 
without his question this letter would never have been born.
The authors also thank Kantaro Ohmori for discussions,
and Shu-Heng Shao and Yunqin Zheng for comments on the draft.
M.O. is supported by FoPM, WINGS Program of the University of Tokyo, 
JSPS Research Fellowship for Young Scientists, JSPS KAKENHI Grant No. JP23KJ0650.
Y.T. is supported by JSPS KAKENHI Grant No. JP24K06883.
Additionally, M.O. and Y.T. are supported in part  
by WPI Initiative, MEXT, Japan at Kavli IPMU, the University of Tokyo.

\bibliographystyle{ytphys}
\bibliography{ref}

\newpage
\appendix
\section{Supplementary materials}

Here we provide additional information which would be useful in understanding the content of the letter.

\subsection{1. More on graphical notation of symmetry actions}
In this section we would like to explain the graphical notation used in the figures in the main part of the paper more concretely, for those who are not very familiar with them.
We demonstrate it using the spin chain language; the continuum description is similar.
We will demonstrate it for more ordinary invertible symmetries.
A concrete non-invertible example is given in the main text.

\def\inc#1{\vcenter{\hbox{\includegraphics[scale=.5]{#1}}}}
Consider a chain of qubits \begin{equation}
\inc{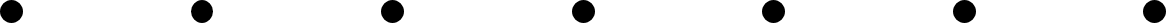}
\end{equation}
where we indicate a qubit by a dot.
Then a local operator $O$ acting on a few consecutive sites can be represented by a blob as follows: \begin{equation}
\inc{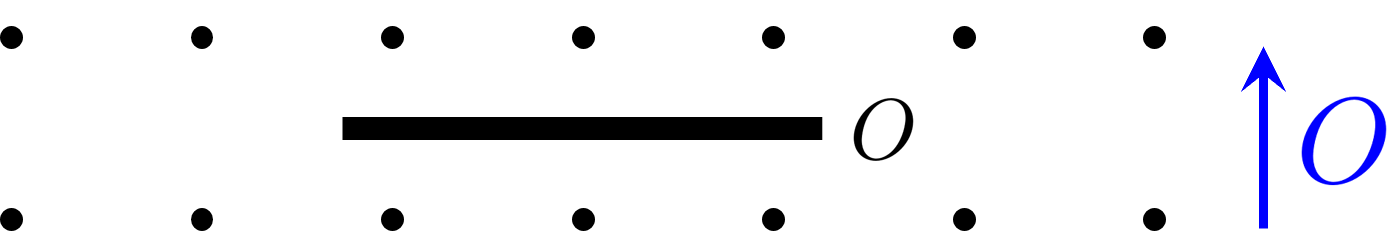}
\end{equation}
where if we consider the bottom row is specifying a ket vector and the top row a bra vector,
then it denotes the matrix element of the operator.
We can also consider the vertical direction as the `time' direction, along which we perform various operations.
The action of $\bZ_2$ symmetry $U_\sg:=\prod_{i} X_i$ on the entire spin chain can be represented similarly as a horizontal line \begin{equation}
\inc{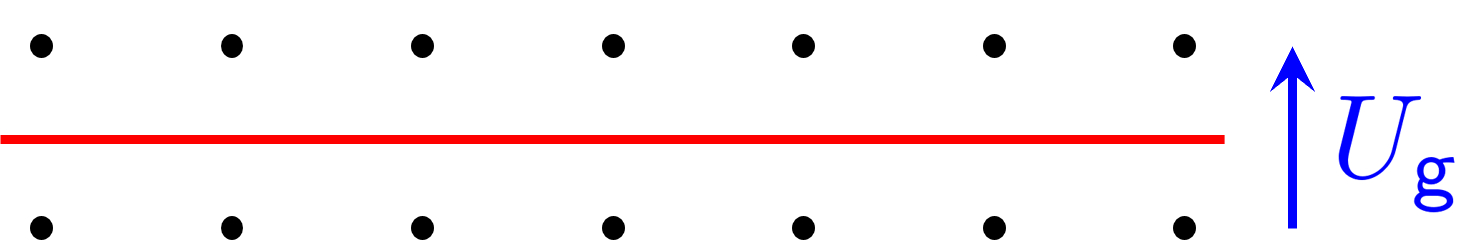}.
\end{equation}

More generally, if the system has a $G$ symmetry and each dot represents a local degree of freedom acted on by $G$, 
two consecutive actions of two symmetry elements $\sg,\sh\in G$ on the entire chain can be denoted as \begin{equation}
\inc{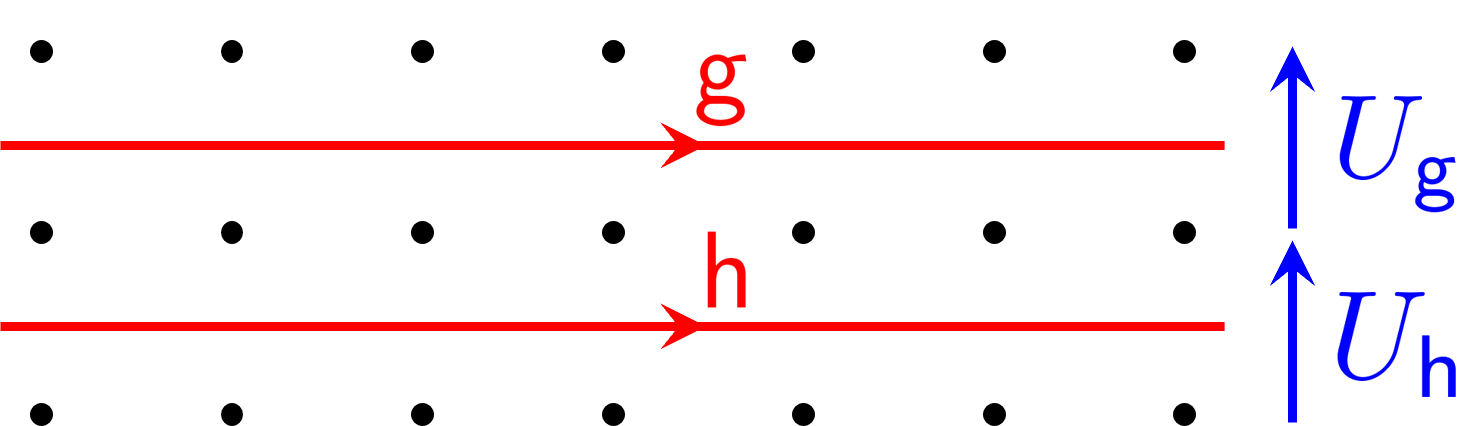}
\end{equation}
which should be equal to the action of $gh\in G$: \begin{equation}
\inc{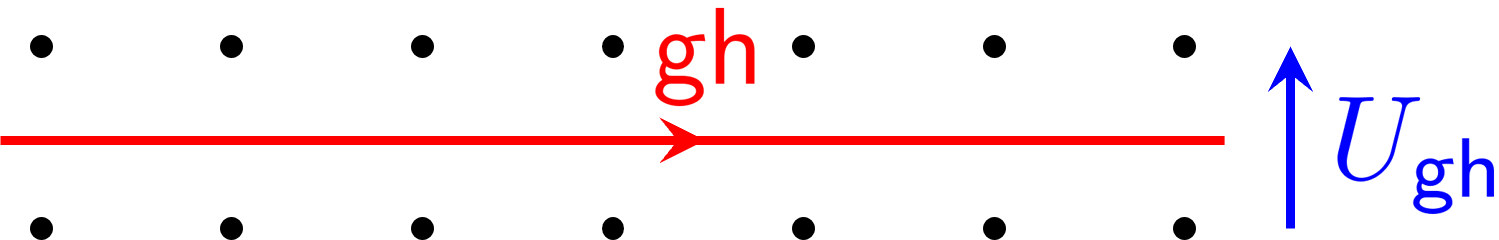}.
\end{equation}
The arrows are attached so that the reversed direction denotes the Hermitian conjugate, say $U_\sg^\dagger$, but omitted when not needed.
\revised{When we are dealing with invertible symmetries,
their actions $U_\sg$ are always unitary,
but when we deal with non-invertible symmetries, like in the main text,
their actions are not unitary in general.}
So, in the main text, we use the notation $A_\sg$ for the action of a symmetry element $\sg$ on the Hilbert space, instead of $U_\sg$.

Coming back to the case of $\bZ_2$ symmetry for simplicity,
we can also consider the action of the $\bZ_2$ symmetry $(U_\sg)_{j\le i\le k}$ only on a finite segment $j\le i\le k$ of the spin chain \begin{equation}
(U_\sg)_{j\le i\le k} = X_j X_{j+1} \cdots X_{k-1} X_k.
\end{equation}
After the action, we have two domain walls, one between $i=j-1$ and $i=j$ and the other between $i=k$ and $i=k+1$, across which we have a boundary condition twisted by $\bZ_2$.
Indeed, if we take the standard transverse Ising model Hamiltonian \begin{equation}
H= \sum_i (Z_i Z_{i+1} + X_i)
\end{equation} and consider $H'=(U_\sg)_{j\le i\le k}^\dagger H (U_\sg)_{j\le i\le k}$, we obtain \begin{equation}
H'= \sum_i ((-1)^{J_i} Z_i Z_{i+1} + X_i) 
\label{H'}
\end{equation}
where $J_i$ is $0$ except at $i=j-1$ and $i=k$.
To encode this effect in the graphical notation, we represent this action in the following manner: 
\begin{equation}
\inc{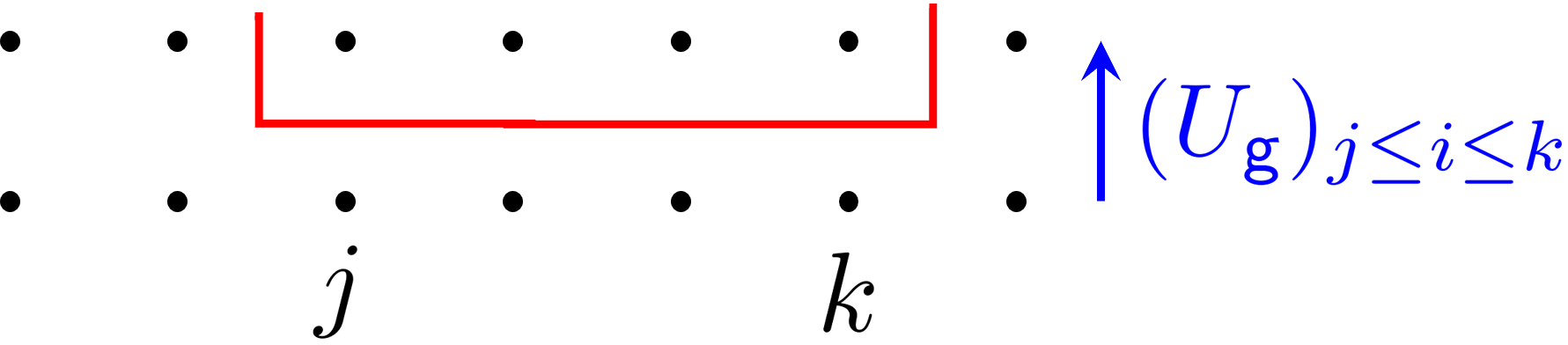}
\end{equation} 
where the Hilbert space of states on the input side and on the output side are distinguished, respectively given by
\begin{equation}
\cH(\cdots \bullet_{j-1} \bullet_{j}\cdots\bullet_k \bullet_{k+1}\cdots)
\end{equation}
and
\begin{equation}
\cH(\cdots \bullet_{j-1} \gwall \bullet_{j}\cdots\bullet_k \gwall   \bullet_{k+1}\cdots),
\end{equation}
and therefore $(U_\sg)_{j\leq i\leq k}$ is considered as a map \begin{multline}
(U_\sg)_{j\leq i\leq k}\colon 
\cH(\cdots \bullet_{j-1}   \bullet_{j}\cdots\bullet_k   \bullet_{k+1}\cdots) \\
\to
\cH(\cdots \bullet_{j-1}   \gwall \bullet_{j}\cdots\bullet_k \gwall   \bullet_{k+1}\cdots).
\end{multline}
Then the action $\cg$ of the $\bZ_2$ symmetry $g$ on a local operator $O$ supported within the segment $j\leq i\leq k$ can be denoted graphically as follows:
\begin{equation}
\vcenter{\hbox{\includegraphics[scale=.45]{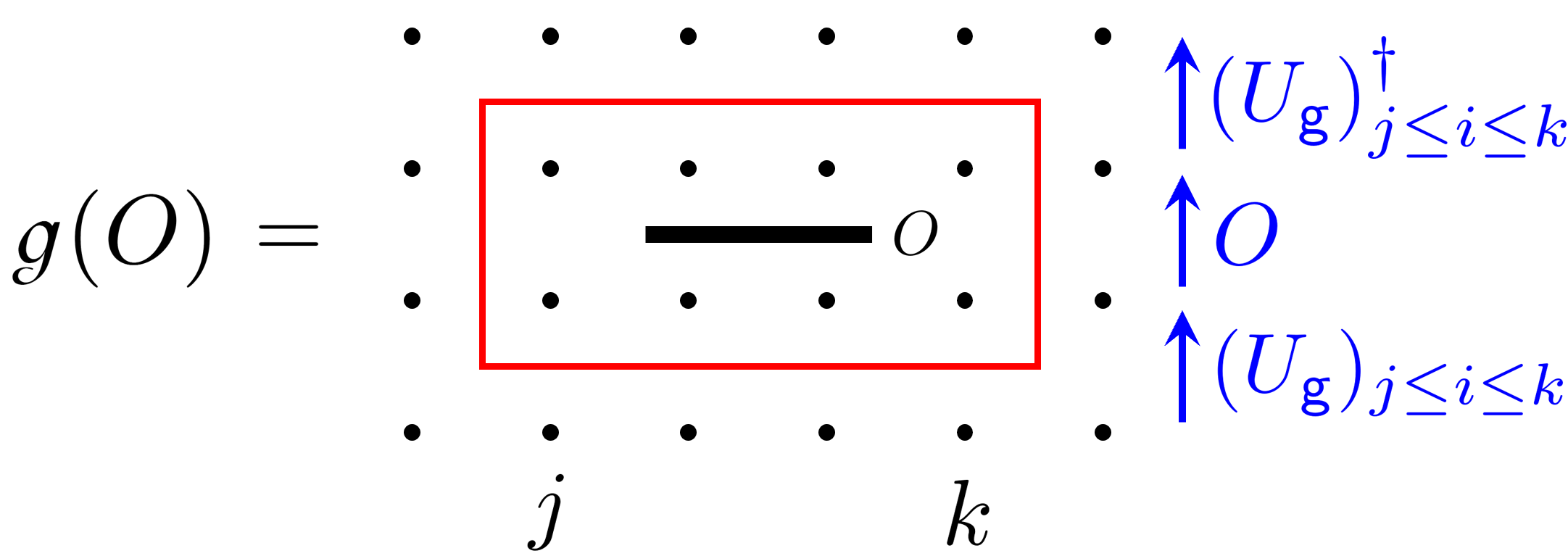}}}
\end{equation} 
Namely, we have \begin{equation}
\cg(O) = (U_\sg)_{j\le i\le k}^\dagger O (U_\sg)_{j\le i\le k}.
\end{equation}

To generalize this to non-invertible symmetries, it turns out convenient to decompose the operation $(U_\sg)_{j\le i \le k}$ into several steps:
\begin{itemize}
\item the introduction of two domain walls between $i=k$ and $i=k+1$,
\item the motion of the wall on the left, step by step, toward the desired point between $i=j-1$ and $i=j$
\item the action of the operator $O$,
\item the reverse motion of the wall on the left back to the point between $i=k$ and $i=k+1$,
\item the removal of the two domain walls.
\end{itemize}
Writing the operator introducing two walls as $V_{\sg\overline{\sg},k}$ 
and the unitary motion of a wall by one step to the right as $R_{\sg,i}$,
we can represent $g(O)$ as follows:
\begin{equation}
\inc{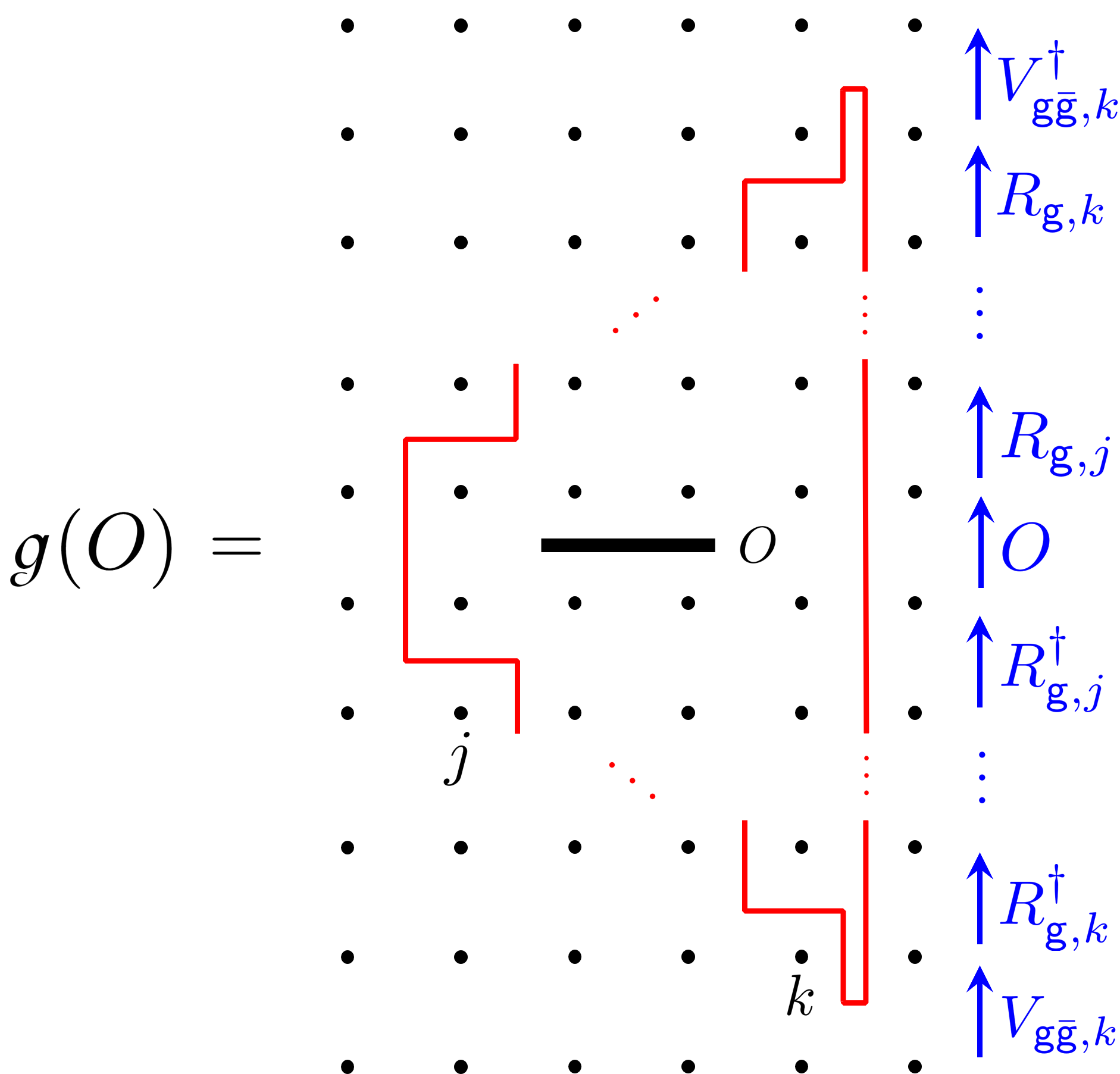}.
\end{equation}
\revised{The generalization of this picture to a non-invertible symmetry, the Kramers--Wannier duality, is shown in Fig.\ \ref{fig:Dwalls} in the main text,
where the representation of the walls may seem a little more complicated
because of the introduction of an ancillary qubit.}

A single domain wall (say between $i=j-1$ and $i=j$) can be created in this manner only by considering the action of the $\bZ_2$ symmetry on the semi-infinite region $-\infty < i \le j-1$ or $j\le i < \infty$.
If we use the former implementation, the position of this wall can be moved in the following manner,
whose meaning should be understandable to a reader who comes to this point:
\begin{equation}
\inc{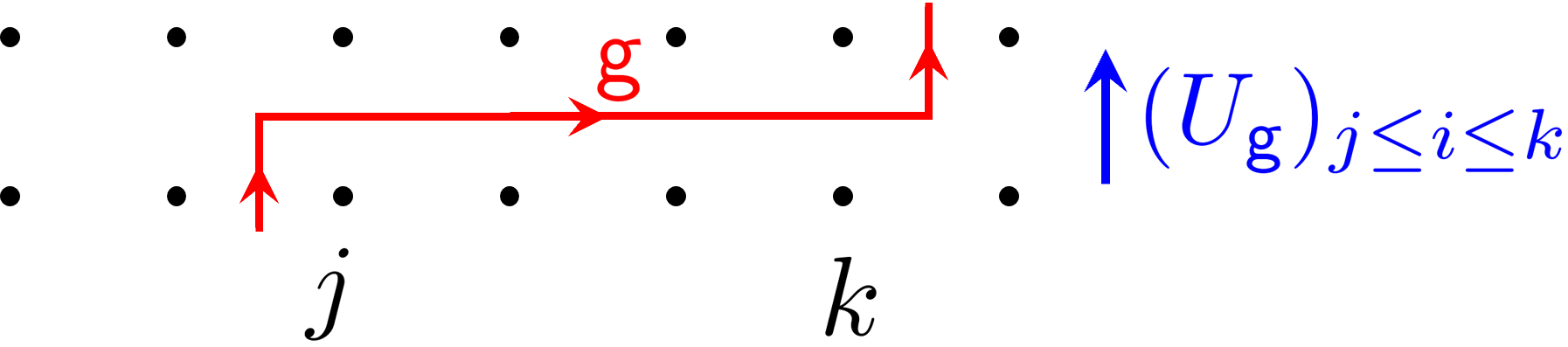}.
\end{equation}

Removing the dots representing the qubits from the figures we introduced above,
we have the version given in Fig.~\ref{fig:walls1} and \ref{fig:walls2}.

\vspace{45pt}

\subsection{2. Explicit Kraus representation of the Kramers--Wannier duality}
Here we convert the expression \eqref{eq:D-action-on-O}
of the local action of the Kramers--Wannier duality,
which was given in the Stinespring representation,
into a Kraus representation
\begin{equation}
\cD(O) = \sum_{t=0,1} K_t O K_t^\dagger.
\end{equation} 
As the method is standard, we simply state the answer, which is given by the following expression:
\begin{widetext}
\begin{align}
\bra{s'_{j-1} s'_j \ldots s'_{k}}K_t\ket{s_{j-1} s_j \ldots s_{k}}&=
\bra{s'_{j-1} s'_j \ldots s'_{k}} V_{\sD\overline{\sD},k}^\dagger R_{\sD,k} R_{\sD,k-1} \cdots R_{\sD,j} \ket{s_{j-1} s_{j} \ldots s_{k} t}\\
&=\frac{1}{\sqrt{2}^{k-j+\frac{3}{2}}} \delta_{s_{j-1}}^{s'_{j-1}} (-1)^{\sum_{i=j}^k( s'_{i-1} s_i + s_i s'_i) + s'_kt}.
\end{align}
\end{widetext}

\end{document}